\documentclass[10pt,a4paper]{article}

\usepackage[T1]{fontenc}
\usepackage[utf8]{inputenc}
\usepackage{lmodern}
\usepackage{textcomp}
\usepackage[a4paper,margin=2.2cm]{geometry}
\usepackage{amsmath,amssymb}
\usepackage{graphicx}
\usepackage{xcolor}
\usepackage{booktabs}
\usepackage{float}
\usepackage{siunitx}
\usepackage[font=small,labelfont=bf,labelsep=period]{caption}
\usepackage{authblk}
\usepackage{titlesec}
\usepackage{enumitem}
\usepackage{microtype}
\usepackage[numbers,sort&compress]{natbib}
\usepackage[colorlinks=true,linkcolor=blue!55!black,citecolor=blue!55!black,urlcolor=blue!55!black]{hyperref}

\graphicspath{{figures/}}
\titleformat{\section}{\Large\bfseries\sffamily}{}{0pt}{}
\titleformat{\subsection}{\large\bfseries\sffamily}{}{0pt}{}
\titleformat{\subsubsection}[runin]{\normalsize\bfseries\sffamily}{}{0pt}{}[.]
\titlespacing*{\section}{0pt}{2.2ex plus .5ex}{0.8ex}
\titlespacing*{\subsection}{0pt}{1.8ex plus .4ex}{0.6ex}
\titlespacing*{\subsubsection}{0pt}{1.2ex plus .3ex}{0.6em}
\newcommand{\Dpar}{D_{\parallel}}
\newcommand{\Dperp}{D_{\perp}}
\newcommand{\Csp}{C_{\mathrm{sp}}}
\newcommand{\dd}{\mathrm{d}}

\newcommand{\flame}{FLAME\textsuperscript{3}-LAB}

\newcommand{\cfmlogo}{%
  \IfFileExists{figures/logo_cfm.png}{\noindent\includegraphics[height=1.4cm]{logo_cfm.png}\par\vspace{0.4em}}{%
  \IfFileExists{figures/logo_cfm.pdf}{\noindent\includegraphics[height=1.4cm]{logo_cfm.pdf}\par\vspace{0.4em}}{}}}

\title{\sffamily\bfseries In-plane diffusion, spurious drift and a physics-structured gap predictor for a free colloidal pair}

\author[1,*]{Juan Pablo Staforelli Vivanco}
\author[1]{Cristobal Augusto Valdés Cruces}
\affil[1]{\flame, Departamento de Física, Facultad de Ciencias Físicas y Matemáticas, Universidad de Concepción, Concepción, Chile}
\affil[*]{Corresponding author: \href{mailto:jstaforelli@udec.cl}{jstaforelli@udec.cl}}

\date{}

\hypersetup{
  pdftitle={In-plane diffusion, spurious drift and a physics-structured gap predictor for a free colloidal pair},
  pdfauthor={Juan Pablo Staforelli Vivanco, Cristobal Augusto Valdés Cruces}}

\begin{document}

\cfmlogo
\maketitle
\vspace{-2.5em}

% ====================================================================
\begin{abstract}
\noindent
When the diffusivity of a Brownian particle depends on position, its mean displacement acquires a noise-induced (spurious) drift whose magnitude is set by the stochastic convention $\alpha$. Inspired by the particle--wall experiments published by G.~Volpe and co-workers, we ask how much of this physics can be extracted from a pair of \SI{1.95}{\micro\metre} polystyrene spheres released from blinking holographic optical tweezers at five temperatures between \SI{2}{\degreeCelsius} and \SI{40}{\degreeCelsius}. The single-sphere diffusivity inside the pair stays between \num{0.093} and \SI{0.098}{\micro\metre\squared\per\second}, whereas the Stokes--Einstein value for water rises from \num{0.126} to \SI{0.361}{\micro\metre\squared\per\second}; an isolated sphere shows the same deficit, so the missing diffusivity scale originates in the measurement and not in the pair interaction. The pair is what makes the diffusivity depend on position: from the measured gap-dependent diffusivities $\Dpar(h)$ and $\Dperp(h)$ we build a Fokker--Planck predictor of the future gap and score it on held-out releases. It recovers the \SI{90}{\percent} width of the gap distribution up to \SI{800}{\milli\second} but cannot discriminate between $\alpha=0$, $1/2$ and $1$, because the $\alpha$-dependent mean shift is an order of magnitude smaller than that width. The ensemble spurious drift is consistent with zero at both ends of the temperature range, and the regression for $\alpha$ never exceeds $R^2=0.22$. We present a planar methods baseline and identify the measurements required to resolve $\alpha$.
\end{abstract}

\noindent\textbf{Keywords:} colloidal pair; blinking optical tweezers; position-dependent diffusion; spurious drift; stochastic convention; trajectory prediction.

% ====================================================================
\section{Introduction}

A Brownian particle whose diffusion coefficient depends on position is described by a stochastic differential equation with multiplicative noise. Such an equation is incomplete until a rule is given for evaluating the noise amplitude within each time step, and the rule is summarised by a single number, the stochastic convention $\alpha\in[0,1]$. Different conventions predict different mean displacements: the difference is a noise-induced contribution, the spurious drift, proportional to $\alpha$ and to the gradient of the diffusivity~\cite{lancon2001,lau2007,volpe2010}. For a particle in thermal equilibrium with its bath, the thermodynamically consistent choice is the isothermal (anti-Itô) value $\alpha=1$~\cite{lau2007,volpe2010}. This was demonstrated for a sphere diffusing in front of a wall by comparing the force obtained from the equilibrium distribution with the force obtained from the drift, once the term $\alpha\,\dd D/\dd z$ had been subtracted~\cite{volpe2010,brettschneider2011}. The same value follows mathematically from the small-mass (Smoluchowski--Kramers) limit of the underlying inertial dynamics, in which the noise-induced drift is fixed by the relation between friction and diffusion: when the two are linked by the Einstein relation, as for a Brownian particle, the limit is the isothermal one, whereas other relations lead to other, possibly position-dependent, values of $\alpha$~\cite{hottovy2012}. The diffusive motion of a colloidal particle close to an interacting surface, with its hydrodynamic coupling to the wall and the first-passage statistics that result, had been treated earlier by Brownian-dynamics simulation~\cite{clark1987}.

The present work is directly inspired by that body of published work by G.~Volpe and his soft-matter group on particle--wall interactions~\cite{volpe2010,brettschneider2011}, and carries its approach over from the sphere--wall geometry to a pair of colloidal spheres. The same mechanism acts on such a pair, in which each sphere plays the role of the wall of the other. Hydrodynamic coupling makes the relative diffusivities along and across the line of centres, $\Dpar$ and $\Dperp$, functions of the surface-to-surface separation~\cite{brenner1961,batchelor1976,crocker1997}. In the focal plane, the projected gap $h=\rho-\sigma$, with $\rho$ the in-plane centre-to-centre distance and $\sigma$ the sphere diameter, obeys in Itô form
\begin{equation}
\dd h=\left\{\alpha\left[\Dpar'(h)+\frac{\Dpar(h)-\Dperp(h)}{\rho}\right]+\frac{\Dperp(h)}{\rho}\right\}\dd t+\sqrt{2\Dpar(h)}\;\dd W ,
\label{eq:sde}
\end{equation}
where the prime denotes a derivative with respect to $h$, $\rho=h+\sigma$ and $W$ is a Wiener process. The term proportional to $\alpha$ is the spurious drift, i.e.\ $\alpha$ times the radial component of the divergence of the diffusion tensor, and $\Dperp/\rho$ is the geometric drift that any radial coordinate acquires in two dimensions. Blinking optical tweezers give direct access to this free dynamics: the particles are positioned by the traps, the traps are switched off, and the subsequent trajectories are recorded without optical forces~\cite{paladugu2016,magazzu2019}. Changing the bath temperature should change $D$ through the viscosity and, in principle, allow a noise-induced drift to be told apart from a force~\cite{volpe2016}.

The usual analysis of such data stops at the conditional moments of Eq.~\eqref{eq:sde}: the local diffusivities and the local drift. Here we go one step further and turn those moments into a predictor. Given the present gap, the predictor returns the full distribution of the gap a time $\tau$ later, together with the probability of first reaching a prescribed separation. Its only inputs are $\Dpar(h)$ and $\Dperp(h)$, calibrated on a training subset of the releases; $\alpha$ enters as a discrete hypothesis and is never fitted. We introduce the predictor because it answers a question that the moments leave open: which part of the stochastic dynamics can be forecast from a single release lasting about one second, and which part is visible only in an ensemble average. We find that the width of the future-gap distribution belongs to the first category and $\alpha$ to the second. Making that separation quantitative on held-out data is the central methodological result of this work.

We state the scope explicitly. This report is a methods baseline intended to improve the measurement of the stochastic convention in a colloidal pair. We do not claim absolute Stokes--Einstein diffusivities for water, and we do not claim a resolved $\alpha(h)$; instead, we document the obstacles that currently prevent such claims, including the limits of our thermal control, and the tests that any future measurement must pass.

% ====================================================================
\section{Results}

\subsection{Experimental platform and predictor pipeline}

\begin{figure}[t]
\centering
\includegraphics[width=0.82\linewidth]{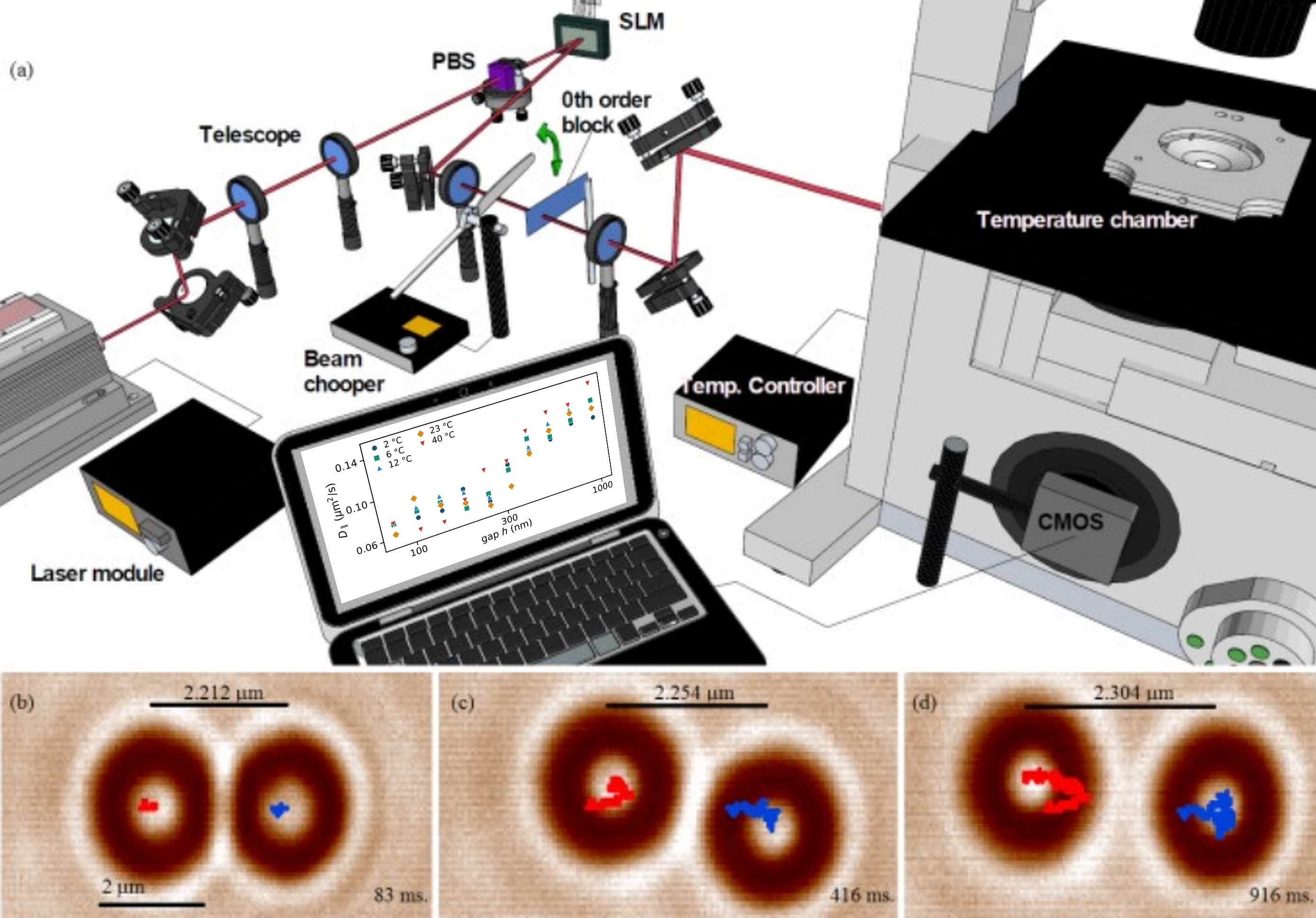}
\caption{\textbf{Blinking holographic optical tweezers.} (\textbf{a}) An expanded, horizontally polarised Nd:YAG laser beam is reflected by a spatial light modulator (SLM) displaying a grey-scale hologram. The conjugate optics image the SLM so that two independent optical traps are generated in the focal (Fourier) plane of the microscope objective. Three-dimensional ($xyz$) video microscopy is performed with a LabVIEW-based acquisition at 600 frames per second using a CMOS camera. The bath temperature is varied from \SI{2}{\degreeCelsius} to \SI{40}{\degreeCelsius}; the set point of the chamber is stabilised with millidegree accuracy by an automatic closed-loop PID controller, but the temperature of the fluid in the focal volume was not measured independently. The computer screen shows, as an illustration, the measured $\Dpar(h)$ of Fig.~\ref{fig:diff}a. (\textbf{b}--\textbf{d}) Three of the 600 frames of one release of the $\sim$\SI{2}{\micro\metre} polystyrene microspheres, taken \SI{83}{\milli\second}, \SI{416}{\milli\second} and \SI{916}{\milli\second} after the traps were switched off by blocking the beam. Red and blue paths are the independent Brownian trajectories of the two spheres in the $xy$ plane.}
\label{fig:setup}
\end{figure}

Figure~\ref{fig:setup} shows the apparatus. An expanded, horizontally polarised Nd:YAG beam is diffracted by a spatial light modulator and relayed to the focal plane of the microscope objective, where it forms two independent traps. A mechanical block of the beam releases the pair, and a CMOS camera records the free motion at 600 frames per second. The bath temperature is set between \SI{2}{\degreeCelsius} and \SI{40}{\degreeCelsius} by a closed-loop controller. The same blinking architecture underlies related pair experiments with holographic tweezers~\cite{paladugu2016,magazzu2019}.

\begin{figure}[t]
\centering
\includegraphics[width=0.95\linewidth]{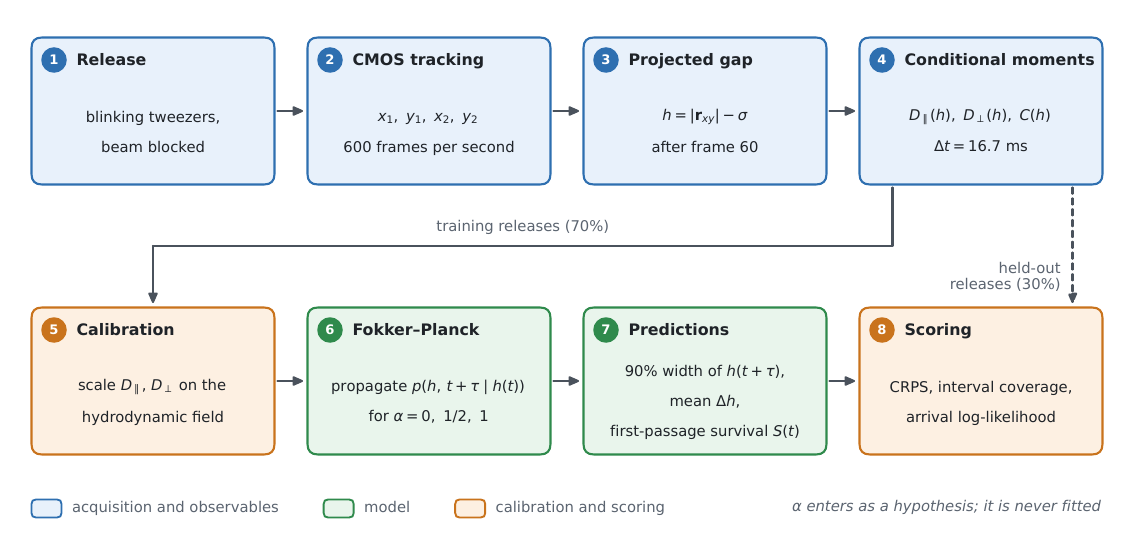}
\caption{\textbf{Operational pipeline of the gap predictor.} Steps 1--4 (blue) are acquisition and observables, steps 6--7 (green) are the model, and steps 5 and 8 (orange) are calibration and scoring. The releases are split once: the training releases (solid branch) calibrate the hydrodynamic field, and the held-out releases (dashed branch) are used only for scoring. The stochastic convention $\alpha$ is never fitted from single-frame increments; it is scored through the ensemble consequences of the propagated density.}
\label{fig:pipeline}
\end{figure}

Figure~\ref{fig:pipeline} summarises the analysis. Each release yields the in-plane positions of the two spheres. The projected gap $h=|\mathbf r_{xy}|-\sigma$ is formed after discarding the first 60 frames ($\approx\SI{100}{\milli\second}$), which removes the shutter and tracker transient. Conditional moments at a sampling interval $\Delta t=\SI{16.7}{\milli\second}$ give $\Dpar(h)$, $\Dperp(h)$ and the radial drift $C(h)$. On a training split of the releases, $\Dpar$ and $\Dperp$ are represented by a hydrodynamic form (Brenner--Faxén or Rotne--Prager, whichever describes the measured profile better) multiplied by a free scale for each component. A small neural network trained on the same releases gives a second, model-free representation of the two diffusivities, used as a cross-check of the field (Methods). With that field fixed, the Fokker--Planck equation associated with Eq.~\eqref{eq:sde} is integrated for $\alpha=0$, $1/2$ and $1$. The outputs are the density $p(h,t+\tau\mid h(t))$, its central \SI{90}{\percent} width, its mean displacement and the first-passage survival probability to a threshold. All scores reported below are computed on held-out releases that did not enter the calibration of $D$.

\subsection{Free trajectories and gap-dependent diffusivities}

\begin{figure}[t]
\centering
\includegraphics[width=0.88\linewidth]{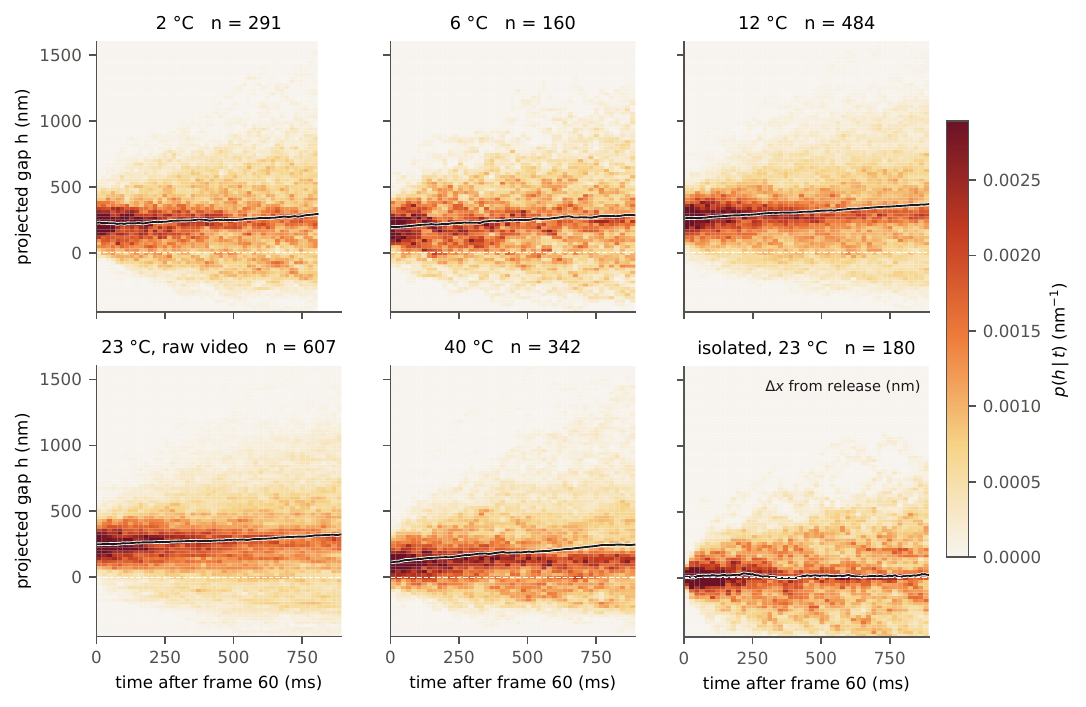}
\caption{\textbf{Projected gap after release, and the isolated sphere.} Colour map of the conditional density $p(h\mid t)$ of the pair at each temperature, with time measured from frame 60; $n$ is the number of releases. The white curve is the ensemble mean and the dashed line is projected contact, $h=0$. The \SI{23}{\degreeCelsius} pair panel is the raw-video measurement. Bottom right: the isolated sphere at \SI{23}{\degreeCelsius} (180 raw videos), shown as the density $p(\Delta x\mid t)$ of its horizontal displacement from the position at frame 60, on the same length, time and colour scales; here the dashed line is the release position, $\Delta x=0$, and the part of the cloud below \SI{-450}{\nano\metre} is cut by the shared axis.}
\label{fig:traj}
\end{figure}

Figure~\ref{fig:traj} shows the evolution of the projected gap after release at the five temperatures. The \SI{23}{\degreeCelsius} panel is measured on the raw camera frames (607 releases); the other panels use the centroids stored by the laboratory acquisition software. At every temperature the ensemble mean moves by less than about \SI{100}{\nano\metre} in \SI{0.8}{\second}, while the cloud spreads by about \SI{1}{\micro\metre}. This disparity anticipates the main result: a predictor based on $D(h)$ can match the width of a single release and still leave $\alpha$ to the ensemble average.

The sampling interval is chosen as the shortest $\Delta t$ at which $\Dpar$ has reached its plateau at every temperature (Supplementary Fig.~\ref{fig:S_interval}), following Magazzù \emph{et al.}~\cite{magazzu2019}. This rule selects ten frames, \SI{16.7}{\milli\second}. A stricter search for a plateau of the drift, referenced to a \SI{75}{\milli\second} lag, does not converge: at \SI{23}{\degreeCelsius} the drift evaluated at \SI{75}{\milli\second} and at \SI{100}{\milli\second} still differ by more than the tolerance. The drift is therefore evaluated at the same \SI{16.7}{\milli\second} interval, at which the short-time transient of the tracker has decayed.

\begin{figure}[t]
\centering
\includegraphics[width=0.92\linewidth]{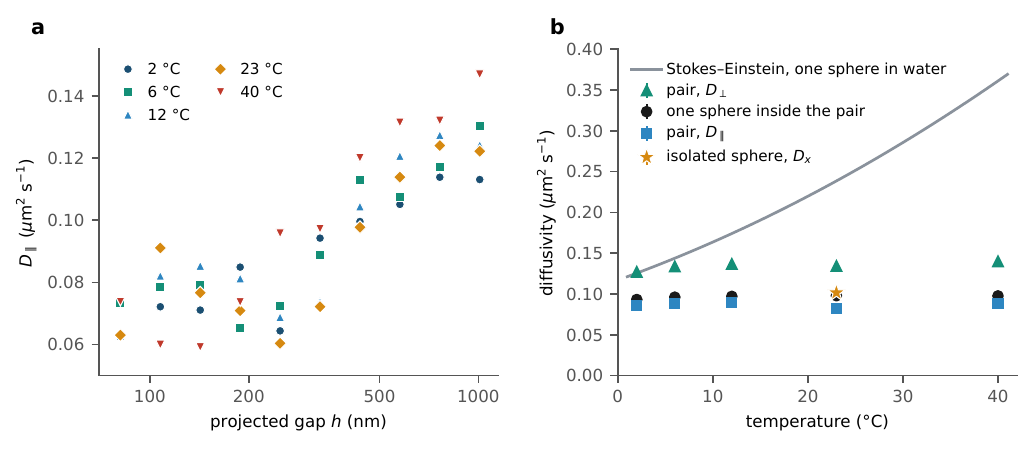}
\caption{\textbf{Diffusivity of the pair.} (\textbf{a}) Conditional $\Dpar(h)$ at the five temperatures; the \SI{23}{\degreeCelsius} points are the raw-video measurement. (\textbf{b}) Diffusivity of one sphere inside the pair and relative diffusivities $\Dpar$ and $\Dperp$ of the pair against temperature, compared with the Stokes--Einstein curve for one sphere in water. The star is $D_x$ of the isolated sphere at \SI{23}{\degreeCelsius}, the component that is reproduced when the raw frames are retracked. Error bars are bootstrap standard deviations and are mostly smaller than the symbols.}
\label{fig:diff}
\end{figure}

Table~\ref{tab:D} and Fig.~\ref{fig:diff} collect the diffusivities. The single-sphere diffusivity inside the pair is \SI{0.093(2)}{\micro\metre\squared\per\second} at \SI{2}{\degreeCelsius} and \SI{0.098(1)}{\micro\metre\squared\per\second} at \SI{40}{\degreeCelsius}. Over the same range the Stokes--Einstein diffusivity of an isolated sphere in water rises from \num{0.126} to \SI{0.361}{\micro\metre\squared\per\second}, a factor of \num{2.87}; the measured value rises by a factor of \num{1.05}. The ordering $\Dperp>\Dpar$ holds at every temperature, as lubrication requires~\cite{brenner1961,batchelor1976}, and $\Dpar(h)$ increases with the gap in the same way in all five baths. Note that $\Dpar$ and $\Dperp$ are diffusivities of the relative coordinate: for two spheres that did not interact, both would equal twice the single-sphere value, $2D\approx\SI{0.19}{\micro\metre\squared\per\second}$. The measured $\Dpar$ and $\Dperp$ are about \SI{45}{\percent} and \SI{70}{\percent} of that value, the hydrodynamic hindrance of the pair at gaps of a few hundred nanometres, which is stronger along the line of centres than across it. Retracking the 607 raw videos at \SI{23}{\degreeCelsius} gives \SI{0.098(2)}{\micro\metre\squared\per\second} for one sphere and $\Dpar=\SI{0.082(2)}{\micro\metre\squared\per\second}$, within \SI{0.01}{\micro\metre\squared\per\second} of the values obtained from the laboratory file for that bath.

\begin{table}[t]
\centering
\caption{\textbf{In-plane diffusivities at $\Delta t=\SI{16.7}{\milli\second}$.} $D$ is the single-sphere diffusivity inside the pair, $D_0$ the Stokes--Einstein value for an isolated sphere in water, and $\Dpar$, $\Dperp$ the relative diffusivities along and across the line of centres. All diffusivities are in \si{\micro\metre\squared\per\second}. Uncertainties are standard deviations over bootstrap resamples of whole releases. The ``video'' row at \SI{23}{\degreeCelsius} is the raw-frame measurement; the ``file'' row is the laboratory record for the same bath.}
\label{tab:D}
\begin{tabular}{lrcccc}
\toprule
Temperature & Releases & $D$ & $D_0$ (water) & $\Dpar$ & $\Dperp$ \\
\midrule
\SI{2}{\degreeCelsius}         & 291 & $0.093\pm0.002$ & 0.126 & $0.086\pm0.003$ & $0.127\pm0.004$ \\
\SI{6}{\degreeCelsius}         & 160 & $0.096\pm0.002$ & 0.144 & $0.088\pm0.003$ & $0.134\pm0.005$ \\
\SI{12}{\degreeCelsius}        & 484 & $0.097\pm0.002$ & 0.174 & $0.089\pm0.002$ & $0.137\pm0.003$ \\
\SI{23}{\degreeCelsius}, file  & 149 & $0.093\pm0.004$ & 0.239 & $0.089\pm0.003$ & $0.133\pm0.006$ \\
\SI{23}{\degreeCelsius}, video & 607 & $0.098\pm0.002$ & 0.239 & $0.082\pm0.002$ & $0.134\pm0.003$ \\
\SI{40}{\degreeCelsius}        & 342 & $0.098\pm0.001$ & 0.361 & $0.088\pm0.003$ & $0.140\pm0.005$ \\
\bottomrule
\end{tabular}
\end{table}

\subsection{Calibration with an isolated sphere}

\begin{figure}[t]
\centering
\includegraphics[width=0.90\linewidth]{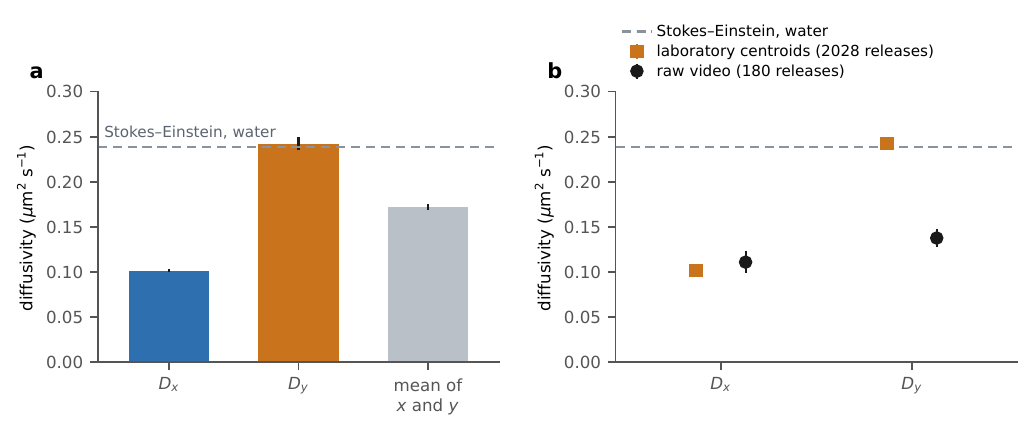}
\caption{\textbf{Isolated sphere at \SI{23}{\degreeCelsius}.} (\textbf{a}) Diffusivities from the laboratory (LabVIEW) centroids, 2028 releases. (\textbf{b}) The same two axes measured on 180 raw videos, next to the laboratory values. The dashed line is the Stokes--Einstein value for water. The horizontal axis agrees between the two measurements and the vertical axis does not, so only $D_x$ is used as the calibration.}
\label{fig:bead}
\end{figure}

To decide whether the missing diffusivity scale originates in the pair, an isolated sphere was recorded in the same \SI{23}{\degreeCelsius} bath, with no second particle, over 2028 releases. On the laboratory centroids, $D_x=\SI{0.102(2)}{\micro\metre\squared\per\second}$ and $D_y=\SI{0.242(7)}{\micro\metre\squared\per\second}$ (Fig.~\ref{fig:bead}); the Stokes--Einstein value at this temperature and radius is \SI{0.239}{\micro\metre\squared\per\second}. Retracking 180 of the raw videos reproduces the horizontal channel, $D_x=\SI{0.111(12)}{\micro\metre\squared\per\second}$, but not the vertical one, $D_y=\SI{0.138(10)}{\micro\metre\squared\per\second}$. The free spreading of this sphere along $x$ is shown in the last panel of Fig.~\ref{fig:traj}: the cloud widens symmetrically about the release position, to a standard deviation of \SI{390}{\nano\metre} after \SI{0.89}{\second}, while its mean, \SI{22}{\nano\metre}, stays within one standard error (\SI{29}{\nano\metre}) of zero. The diffusivity confirmed by the images is therefore the horizontal value, about \num{0.43} of the Stokes--Einstein number, and the pair diffusivities in Table~\ref{tab:D} lie on that same scale.

Two conclusions follow. First, the deficit with respect to water is already present for one sphere, so it is not generated by the hydrodynamic interaction of the pair. Second, a single rescaling of the pixel size cannot remove it: the factor that matches $D_x$ to water would make the laboratory $D_y$ about \num{2.4} times too large.

\subsection{What a single release can forecast}

\begin{figure}[t]
\centering
\includegraphics[width=0.92\linewidth]{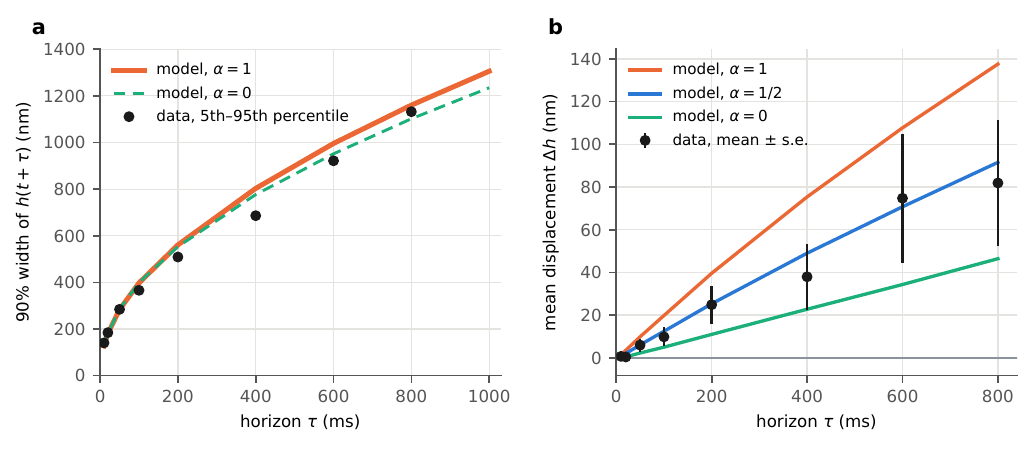}
\caption{\textbf{Gap predictor on held-out releases at \SI{12}{\degreeCelsius}.} (\textbf{a}) Central \SI{90}{\percent} width of $h(t+\tau)$ against the horizon $\tau$, for the data (points) and for the model with $\alpha=1$ (solid) and $\alpha=0$ (dashed). The two model curves coincide below about \SI{200}{\milli\second} and differ by \SI{5}{\percent} at \SI{800}{\milli\second}; the model is evaluated up to \SI{1}{\second}, beyond the last horizon with data. (\textbf{b}) Ensemble mean displacement against the horizon, with its standard error, and the model curves for $\alpha=0$, $1/2$ and $1$. The width follows the calibrated $D(h)$; the mean lies between the conventions.}
\label{fig:pred}
\end{figure}

On held-out releases at \SI{12}{\degreeCelsius}, starting from a window of $\pm\SI{35}{\nano\metre}$ about the median gap, the measured \SI{90}{\percent} interval of $h(t+\tau)$ is \SI{141}{\nano\metre} wide at $\tau=\SI{10}{\milli\second}$ and \SI{1132}{\nano\metre} wide at \SI{800}{\milli\second}. The predictor gives \SI{128}{\nano\metre} and \SI{1160}{\nano\metre} (Fig.~\ref{fig:pred}a; Supplementary Table~\ref{tab:S_scores}). The agreement is not uniform across horizons: between \SI{100}{\milli\second} and \SI{400}{\milli\second} the model overestimates the measured width by about \SIrange{8}{13}{\percent}, the largest difference being at \SI{400}{\milli\second}, where the data give \SI{686}{\nano\metre} and the model \SI{777}{\nano\metre} ($\alpha=0$) and \SI{803}{\nano\metre} ($\alpha=1$). The continuous ranked probability score (CRPS) at \SI{800}{\milli\second} is \SI{201}{\nano\metre} for a local Brownian baseline, \SI{199}{\nano\metre} for $\alpha=0$ and \SI{199}{\nano\metre} for $\alpha=1$. Knowing $\alpha$ therefore does not improve the distributional forecast of a single trajectory. What the forecast requires is $D(h)$: without the calibrated field, the width would be wrong by hundreds of nanometres at long horizons. The same holds at every horizon from \SI{10}{\milli\second} to \SI{800}{\milli\second} (Supplementary Table~\ref{tab:S_full12}): the width of the prediction is set by $\Dpar(h)$, which is common to the three values of $\alpha$, and $\alpha$ only moves the mean.

The mean displacement of the same held-out sample, now using every test release with $h(t)$ between \SI{150}{\nano\metre} and \SI{500}{\nano\metre}, is \SI{82(30)}{\nano\metre} at \SI{800}{\milli\second} (Fig.~\ref{fig:pred}b). The two extreme conventions predict \SI{46}{\nano\metre} ($\alpha=0$) and \SI{138}{\nano\metre} ($\alpha=1$), with \SI{92}{\nano\metre} for $\alpha=1/2$, and the datum lies between them. The $\alpha$-dependent shift is thus of order \SI{100}{\nano\metre}, whereas the \SI{90}{\percent} width is of order \SI{1100}{\nano\metre}. This ratio is the quantitative content of the statement that a trajectory is anticipated by its width: the predictor anticipates the interval that will contain the future gap, not the point inside that interval that a given release will occupy.

At \SI{23}{\degreeCelsius} the three conventions begin to separate in the score (Supplementary Table~\ref{tab:S_full23}): from \SI{200}{\milli\second} onwards $\alpha=1$ lowers the CRPS below the Brownian baseline, a skill of about 2 to 6 percentage points, because the accumulated spurious drift is no longer small compared with the diffusive width. The test sample is small, however (45 releases, and 30 starting points at \SI{800}{\milli\second}), and the uncertainty of the measured displacement, \SI{98(60)}{\nano\metre} at \SI{800}{\milli\second}, is of the same order as the difference between conventions, so this is an indication and not a ranking.

The first-passage time from $h_0\approx\SI{280}{\nano\metre}$ to \SI{500}{\nano\metre} separates the conventions inside the model, with mean arrival times of \SI{1140}{\milli\second} and \SI{753}{\milli\second} for $\alpha=0$ and $\alpha=1$ at \SI{12}{\degreeCelsius}, but not in the records: 65 of 128 test releases arrive within the video, and the censored mean log-likelihoods are $-0.48\pm0.05$ ($\alpha=0$) and $-0.52\pm0.06$ ($\alpha=1$) (Supplementary Table~\ref{tab:S_fp}). A synthetic control with the same $D$, no tracker and a known $\alpha$ assigns the higher likelihood to the true convention by about \num{0.03} nats (Supplementary Information). The test therefore has the correct sign, but a one-second video is too short to resolve it.

\subsection{Ensemble spurious drift and the effective convention}

\begin{figure}[t]
\centering
\includegraphics[width=0.92\linewidth]{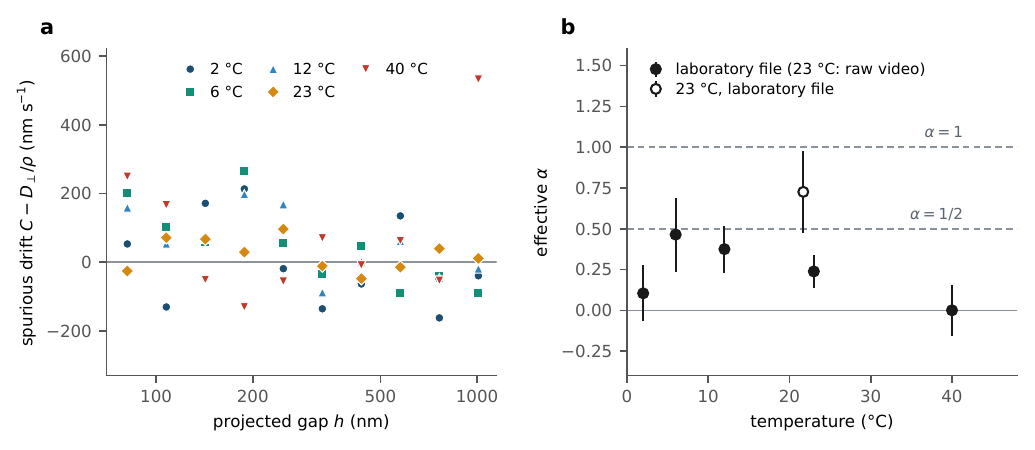}
\caption{\textbf{Spurious drift from the ensemble.} (\textbf{a}) Model-free spurious drift $C-\Dperp/\rho$ against the projected gap; the \SI{23}{\degreeCelsius} points are the raw-video measurement. (\textbf{b}) Effective $\alpha$ at each temperature, with its bootstrap uncertainty; at \SI{23}{\degreeCelsius} the filled symbol is the raw-video value and the open symbol the value from the laboratory file. Dashed lines mark $\alpha=1/2$ and $\alpha=1$. The coefficients of determination are listed in Table~\ref{tab:drift}.}
\label{fig:drift}
\end{figure}

The quantity that carries $\alpha$ is the ensemble radial drift. Subtracting the measured geometric term $\Dperp/\rho$ from the conditional radial velocity $C(h)$ gives a model-free estimate of the spurious drift, $\Csp$, listed in Table~\ref{tab:drift}. At \SI{2}{\degreeCelsius} and \SI{40}{\degreeCelsius} it is consistent with zero. At \SI{12}{\degreeCelsius} it is \SI{41(26)}{\nano\metre\per\second} and the effective convention is $\alpha=0.38\pm0.14$ with $R^2=0.12$. No temperature returns $R^2$ above \num{0.22}, so the dependence $\alpha(h)$ is not resolved: the quoted $\alpha$ is a ratio of two averages. Figure~\ref{fig:drift} shows the bin-by-bin scatter.

On the raw frames at \SI{23}{\degreeCelsius} the same estimator gives \SI{22(16)}{\nano\metre\per\second} and $\alpha=0.24\pm0.10$ ($R^2=0.18$), smaller than the \SI{86(42)}{\nano\metre\per\second} obtained from the laboratory file for that bath. The two centroid definitions share the fast motion but not the sub-hertz baseline (Supplementary Information), so the drift, unlike $D$, depends on which centroid is used.

The predictor and the ensemble drift are therefore complementary. The former states what one release can support; the latter is the estimator through which $\alpha$ has to be sought, and with five temperatures it remains undetermined.

\begin{table}[t]
\centering
\caption{\textbf{Spurious drift and effective convention at $\Delta t=\SI{16.7}{\milli\second}$.} $\Csp$ is the model-free spurious drift $C-\Dperp/\rho$, and $R^2$ is the fraction of its variance explained by $\alpha$ times the hydrodynamic divergence (Supplementary Eqs.~\ref{eq:S_xy} and~\ref{eq:S_alpha}).}
\label{tab:drift}
\begin{tabular}{lrcc}
\toprule
Temperature & $\Csp$ (\si{\nano\metre\per\second}) & $\alpha$ & $R^2$ \\
\midrule
\SI{2}{\degreeCelsius}         & $2\pm31$  & $0.10\pm0.17$ & 0.02 \\
\SI{6}{\degreeCelsius}         & $47\pm40$ & $0.46\pm0.23$ & 0.22 \\
\SI{12}{\degreeCelsius}        & $41\pm26$ & $0.38\pm0.14$ & 0.12 \\
\SI{23}{\degreeCelsius}, file  & $86\pm42$ & $0.73\pm0.25$ & 0.21 \\
\SI{23}{\degreeCelsius}, video & $22\pm16$ & $0.24\pm0.10$ & 0.18 \\
\SI{40}{\degreeCelsius}        & $-6\pm39$ & $0.00\pm0.15$ & 0.00 \\
\bottomrule
\end{tabular}
\end{table}

% ====================================================================
\section{Discussion}

The predictor introduced here is not a device for guessing the next centroid of a sphere. It is a Fokker--Planck map from the present gap to the distribution of the future gap, built from the same hydrodynamic field that enters the spurious-drift analysis. Its role is to make that map quantitative on held-out data and to show that it is informative about the width of the distribution and uninformative about $\alpha$. Without it, the statement that ``$\alpha$ is an ensemble quantity'' would be an order-of-magnitude remark. With it, the remark becomes a measured ratio: at \SI{800}{\milli\second} the $\alpha$-dependent mean shift is an order of magnitude smaller than both the predicted and the observed width, and the distributional score does not prefer one convention.

This result invites a reflection on the curious status of $\alpha$. The convention is not a property of any single trajectory. For a Brownian path the diffusivity can in principle be read from one realisation, through the quadratic variation of its displacements, but the drift cannot: it emerges only from an average, over many releases or over many returns to the same position. Since $\alpha$ has measurable content only through the noise-induced drift, once forces are excluded and $D(h)$ is known, it belongs to the probability law of the displacements and not to an individual realisation of the motion. The situation is reminiscent of the variational principles of physics. A projectile obeys the principle of least action, which is formulated over the whole family of conceivable trajectories, yet it travels only one of them. A light ray crossing between two media follows the path of least time (Fermat's principle), a global condition that seems to require knowledge of the end point, which the ray leaving the starting point does not have. In both cases a statement about many paths governs a motion that takes place only once. The resemblance is more than formal: in the path-integral formulation of state-dependent diffusion the transition probability is a sum over all paths, and $\alpha$ enters the weight of each one~\cite{lau2007}. The difference is that here the family of paths can be sampled, by repeating the release under nominally identical conditions. A single trajectory could replace the ensemble only if it revisited the same gap many times, which a free pair that diffuses apart within a second does not do.

Three limitations stand between this baseline and a claim about $\alpha$; the first two concern the analysis and the third the experiment. The first concerns the diffusivity scale. $D$ does not follow water from \SI{2}{\degreeCelsius} to \SI{40}{\degreeCelsius}, and the isolated sphere shows that the deficit is already present, and anisotropic, for one particle. Until the viscosity of the samples and the origin of the difference between $D_x$ and $D_y$ are established, the hydrodynamic divergence in the denominator of $\alpha$ inherits that ambiguity. The second concerns the drift itself. $R^2$ of the $\alpha$ regression is low at every temperature, and the raw-video centroid does not return the same drift as the laboratory file. We also tested, rather than assumed, the planar description: the radius of the dark diffraction ring, the only defocus cue available in the frames, does not reach a plateau with lag (Supplementary Fig.~\ref{fig:S_axial}) and is distorted by the neighbouring sphere at small gaps (Supplementary Fig.~\ref{fig:S_rings}), so it does not qualify as an axial coordinate and Eq.~\eqref{eq:sde} remains planar.

A third limitation, and possibly the main one, is thermal control. The controller stabilises the temperature of the chamber, but the temperature of the fluid in the focal volume was not measured independently, and the immersion oil is a likely source of artefacts in two ways. First, the oil is in direct contact with the sample cell and, together with the objective lens, can act as a heat sink for the thermal bath, so that the observed region need not follow the set point; a diffusivity that stays nearly flat between \SI{2}{\degreeCelsius} and \SI{40}{\degreeCelsius} is compatible with such a thermal leak. Second, the oil is specified for operation near room temperature, so that imaging artefacts may appear towards the ends of the temperature range explored here, which lie outside the range of the oil. Both effects justify reducing the magnification optics so as to operate without immersion, with only air between the objective and the sample cell. The price is a lower magnification and resolution, which demands additional, non-invasive resources for enhancing resolution and tracking; we return to them below.

These limitations define the measurements that should come next. On the diffusivity scale, the viscosity of the suspension should be measured from \SI{2}{\degreeCelsius} to \SI{40}{\degreeCelsius}, together with the temperature in the focal volume during a release, so that the comparison with $T/\eta$ is made on the fluid that is actually in the cell. The reason why the laboratory $D_y$ of the isolated sphere matches water while $D_x$ and the raw-video $D_y$ do not must also be found: a square pixel and a round diffraction ring cannot absorb a factor of \num{2.4} between the two axes.

On the drift, the number of releases should be increased until the uncertainty of the spurious drift falls well below its value for one centroid definition, and the raw-video and laboratory centroids should be required to return the same drift before any $\alpha$ is quoted. The divergence should be reconstructed from a smoothed measured $\Dpar(h)$ as well as from the hydrodynamic field, and a positive coefficient of determination should be required before quoting $\alpha(h)$. Adding salt to set a Debye length of a few nanometres would remove electrostatic forces from the window $h>\SI{100}{\nano\metre}$.

The quality of the recorded positions is itself a limiting factor, as the disagreement between the two centroid definitions shows, and it would become more so under the immersion-free, lower-magnification regime proposed above. Instrumental improvements in data recording are therefore a natural route forward. Sub-pixel motion retrieved from high-speed, low-resolution video sequences can be used to apply superresolution to images of optically manipulated particles~\cite{staforelli2010}, and event-based (neuromorphic) cameras assisted by optical modulation of the image increase the number of recorded events and improve the tracking of Brownian microparticles~\cite{huenchual2025}. Both approaches aim at improving the records of motion (particle tracking) by purely imaging means, that is, without introducing invasive detection forces on the particle. Either one could raise the temporal resolution and the length of the free trajectories well beyond the present 600 frames per release.

Finally, future work should aim to characterise the phenomenon in three dimensions. The sphere--wall experiments that inspired this study resolve the coordinate normal to the wall, whereas the present baseline is restricted to the projection of the pair onto the focal plane. A three-dimensional description, with the full centre-to-centre separation in place of the projected gap, requires an axial coordinate calibrated against a known displacement; any candidate must first pass the plateau test used here for $x$. Meanwhile, the predictor remains useful as a validation of $D(h)$ on held-out widths, and first-passage ranking of $\alpha$ becomes meaningful only when the video is longer than the separation between the mean arrival times predicted for $\alpha=0$ and $\alpha=1$.

% ====================================================================
\section{Methods}

\subsection{Experimental setup and acquisition}

Polystyrene spheres of diameter $\sigma=\SI{1.95}{\micro\metre}$ were trapped by two holographic foci (Fig.~\ref{fig:setup}) and released by blocking the laser beam. The free motion was recorded at a median frame interval of \SI{1.672}{\milli\second} (2, 6, 12 and \SI{40}{\degreeCelsius}) and \SI{1.675}{\milli\second} (\SI{23}{\degreeCelsius}). Releases in which the images of the two spheres fused were removed. The relative vector is $\mathbf r=\mathbf x_2-\mathbf x_1$ in the focal plane. The isolated-sphere calibration uses the 2028 laboratory files at \SI{23}{\degreeCelsius} and a uniform sample of 180 raw videos.

\subsection{Particle tracking}

Lateral positions in the laboratory files are grey-level centroids at a nominal scale of \SI{49}{\nano\metre} per pixel. At \SI{23}{\degreeCelsius} the raw frames, which are a crop around the spheres rather than the full sensor, were tracked again. Each sphere appears as a bright core inside a dark ring; the reported centre is the intensity-weighted centroid of that core, the class of estimator reviewed by Helgadottir, Argun and Volpe~\cite{helgadottir2019}. On the 117 releases for which both a video and a laboratory file exist, the correlation between the two measurements of the projected gap, after high-pass filtering (moving-average window of 41 frames, about \SI{70}{\milli\second}), has a median of \num{0.965} and a tenth percentile of \num{0.950}. The correlation of the unfiltered gap has a median of \num{-0.06}: the two centroids share the Brownian increments but not the sub-hertz baseline.

\subsection{Conditional moments}

After frame 60, trajectories were subsampled every $n$ frames. The increments of $\mathbf r$ over one sampling interval $\Delta t$ were projected onto the directions parallel and perpendicular to the line of centres at the previous subsampled time, and binned in $h$. In each bin, the diffusivities $\Dpar(h)$ and $\Dperp(h)$ were obtained from the second moments of the projected increments with the covariance-based estimator of Vestergaard \emph{et al.}~\cite{vestergaard2014}, and the radial drift from the first moment, $C(h)=\langle\Delta h\mid h\rangle/\Delta t$. The plateau rule described in the Results selected $n=10$. Uncertainties are standard deviations over bootstrap resamples of whole releases.

\subsection{Reference diffusivity and hydrodynamic field}

The reference for an isolated sphere is the Stokes--Einstein diffusivity $D_0=k_{\mathrm B}T/(3\pi\eta\sigma)$, with $\eta$ the viscosity of water at the bath temperature $T$. Two hydrodynamic forms were considered for the gap dependence of the pair diffusivities: the Brenner--Faxén form~\cite{brenner1961,derjaguin1934} and the Rotne--Prager tensor~\cite{rotne1969}. On the training releases the Kramers--Moyal profile of $\Dpar$ and $\Dperp$ was estimated in bins of $h$, and a free scale, different for each component, was fitted by weighted least squares to each form over $\SI{80}{\nano\metre}\le h\le\SI{950}{\nano\metre}$. The form with the larger sum of the coefficients of determination of $\Dpar$ and $\Dperp$ was retained: Rotne--Prager at \SI{12}{\degreeCelsius} and Brenner--Faxén at \SI{23}{\degreeCelsius} (Supplementary Table~\ref{tab:S_calib}). The convention $\alpha$ does not enter this fit. The fitted scales lie between \num{0.45} and \num{0.66}; a scale below one absorbs the effective viscosity of the sample and the smoothing introduced by the tracker, so the predictor does not rely on the absolute diffusivity of water. The spurious drift and $\alpha$ were obtained with the occupancy-weighted regression described in the Supplementary Information.

\subsection{Neural-network representation of the diffusivities}

As an alternative to the hydrodynamic form, the two diffusivities were also represented by a neural network. The network is a multilayer perceptron with three hidden layers of 64 units and SiLU activation. Its input is $\log h$ and its two outputs, passed through a softplus function so that they remain positive, are $\Dpar(h)$ and $\Dperp(h)$. The derivative $\Dpar'(h)$ that enters the drift of Eq.~\eqref{eq:sde} is obtained by propagating the tangent through the network (forward-mode differentiation), not by finite differences. The network was trained on the individual increments of the training releases at the sampling interval $\Delta t$, for gaps between \SI{30}{\nano\metre} and \SI{1.2}{\micro\metre} and after removing the \SI{0.5}{\percent} largest targets. For each increment, covariance-estimator targets $y_\parallel$ and $y_\perp$ were formed from two consecutive displacements projected on the line of centres and across it~\cite{vestergaard2014}; their expectations are $2\Dpar\Delta t$ and $2\Dperp\Delta t$ and are unbiased by localisation noise. The loss is the sum of two squared errors, each divided by the variance of its target,
\begin{equation}
\mathcal{L}=\frac{\big\langle (y_\parallel-2\Dpar\Delta t)^2\big\rangle}{\mathrm{Var}(y_\parallel)}+\frac{\big\langle (y_\perp-2\Dperp\Delta t)^2\big\rangle}{\mathrm{Var}(y_\perp)} ,
\label{eq:loss}
\end{equation}
minimised with the Adam optimiser (learning rate \num{3e-3}, halved after ten epochs without improvement; mini-batches of 4096 samples; at most 40 epochs), with \SI{15}{\percent} of the samples set aside for validation and early stopping. The network saw \num{14665} samples at \SI{12}{\degreeCelsius} (final validation loss \num{2.08}) and \num{4556} at \SI{23}{\degreeCelsius} (\num{1.84}). A loss close to two is expected: each target is a single noisy squared increment, so a smooth $D(h)$ can explain only a small part of its variance.

The architecture is designed so that the same network can also assemble a drift with the physical structure of Eq.~\eqref{eq:sde}, $C=\alpha X+G$ with $X$ the radial divergence of the diffusion tensor and $G$ the geometric term, and solve $\alpha$ in closed form at every training step, because the drift is linear in $\alpha$. In the present predictor only the loss of Eq.~\eqref{eq:loss} is trained, so $\alpha$ is not learnt: the single-step drift does not inform $\alpha$ ($R^2\le 0$), and the convention is judged by the long-horizon predictions instead. The network curve is compared with the scaled hydrodynamic form and with the measured profile in Supplementary Fig.~\ref{fig:S_net}. It is not used for the scores: the density is always propagated with the scaled hydrodynamic field.

\subsection{Predictor and scoring}

With $D(h)$ fixed, the density of the gap evolves according to the Fokker--Planck equation equivalent to Eq.~\eqref{eq:sde},
\begin{equation}
\frac{\partial p}{\partial t}=-\frac{\partial}{\partial h}\Big[v_\alpha(h)\,p\Big]+\frac{\partial^2}{\partial h^2}\Big[\Dpar(h)\,p\Big],
\qquad
v_\alpha(h)=\alpha\left[\Dpar'+\frac{\Dpar-\Dperp}{\rho}\right]+\frac{\Dperp}{\rho},
\label{eq:fp}
\end{equation}
which was integrated with a Crank--Nicolson scheme on a grid with reflecting boundaries for free evolution, and with an absorbing boundary at \SI{500}{\nano\metre} for first passage. Numerical parameters are listed in the Supplementary Information. Releases were split at random into training (\SI{70}{\percent}) and test sets. The scores on the test set were the continuous ranked probability score, $\mathrm{CRPS}(F,y)=\int[F(z)-\mathbf 1(z\ge y)]^2\,\dd z$ for a predicted cumulative distribution $F$ and an observed gap $y$; the width and coverage of the central \SI{90}{\percent} interval; and the censored log-likelihood of first arrival. For each horizon $\tau$, starting points were taken on the test releases with $h_0$ between \SI{150}{\nano\metre} and \SI{500}{\nano\metre}, each separated from the previous one by at least $\tau$ (and never by less than \SI{80}{\milli\second}). The measured displacement is the mean of $h(\tau)-h_0$, with the standard error from a bootstrap over releases; the model displacement is the mean of the density propagated from the same $h_0$ for $\alpha=0$, $1/2$ and $1$. Because $\Dpar$ is the same in the three cases, the difference between them is only the accumulated spurious drift. The local Brownian baseline uses $\Dpar(h_0)$ and no drift: $h(\tau)$ is compared with a normal distribution of mean $h_0$ and variance $2\Dpar(h_0)\tau$. At $\tau=\SI{1.00}{\second}$ there is no empirical pair, because the free window after frame 60 ends at about \SI{0.89}{\second}; for that horizon only the model is reported, started from the median gap at release (\SI{266}{\nano\metre} at \SI{12}{\degreeCelsius} and \SI{250}{\nano\metre} at \SI{23}{\degreeCelsius}). A fit of $\alpha$ to single-frame increments was also computed and is not used for prediction: it gives $0.10\pm0.37$ ($R^2=-0.22$) at \SI{12}{\degreeCelsius} and $-0.28\pm0.21$ ($R^2=-0.32$) at \SI{23}{\degreeCelsius}, which is not a measurement.

% ====================================================================
\section{Data availability}

The laboratory records and the analysis code that support the findings of this study are archived at \flame{} and are available from the corresponding author upon reasonable request.

% ====================================================================

% ====================================================================
\section{Acknowledgements}

The authors thank Dr.~Giovanni Volpe for conceiving the original idea and establishing the theoretical basis of this work, and Dr.~Agnese Callegari for the initial data treatment, interpretation and analysis codes. Both are members of the Soft Matter Lab (Volpe Lab) at the University of Gothenburg (\url{https://softmatterlab.org/volpe-lab/}).

\section{Author contributions}

J.P.S.V. designed the measurement programme and supervised the experiment. C.A.V.C. carried out the acquisition. Both authors designed the predictor pipeline, interpreted the results and wrote the manuscript.

\section{Competing interests}

The authors declare no competing interests.

% ====================================================================
%  SUPPLEMENTARY INFORMATION
% ====================================================================
\clearpage
\setcounter{figure}{0}\renewcommand{\thefigure}{S\arabic{figure}}
\setcounter{table}{0}\renewcommand{\thetable}{S\arabic{table}}
\setcounter{equation}{0}\renewcommand{\theequation}{S\arabic{equation}}
\renewcommand{\theHfigure}{S\arabic{figure}}
\renewcommand{\theHtable}{S\arabic{table}}
\renewcommand{\theHequation}{S\arabic{equation}}

\begin{center}
{\LARGE\sffamily\bfseries Supplementary Information}\\[0.6em]
{\large\sffamily In-plane diffusion, spurious drift and a physics-structured gap predictor for a free colloidal pair}\\[0.4em]
J. P. Staforelli Vivanco and C. A. Valdés Cruces
\end{center}

\subsection{S1. Sampling-interval plateau}

\begin{figure}[H]
\centering
\includegraphics[width=0.55\linewidth]{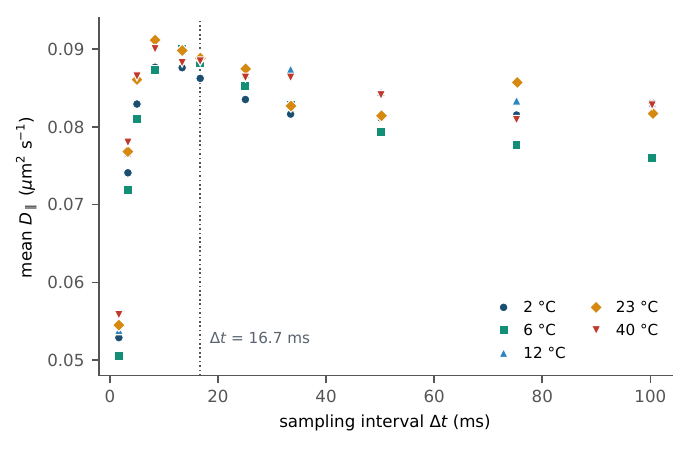}
\caption{\textbf{Choice of sampling interval.} Occupancy-weighted mean of $\Dpar$ against the sampling interval $\Delta t$ at the five temperatures. The dotted line is ten frames, \SI{16.7}{\milli\second}.}
\label{fig:S_interval}
\end{figure}

Supplementary Fig.~\ref{fig:S_interval} shows the occupancy-weighted mean of $\Dpar$ against $\Delta t$ at the five temperatures. The short-time transient is over by about \SI{10}{\milli\second}; the dotted line is the interval adopted for both diffusion and drift, ten frames.

\subsection{S2. Predictor scores}

Supplementary Tables~\ref{tab:S_scores}--\ref{tab:S_full23} collect the held-out scores. At \SI{800}{\milli\second} and \SI{12}{\degreeCelsius} the CRPS is \SI{201}{\nano\metre} for the local Brownian baseline and \SI{199}{\nano\metre} for both $\alpha=0$ and $\alpha=1$, i.e.\ independent of $\alpha$ at the reported precision, while the width follows the calibrated $D$ (\SI{1102}{\nano\metre} for $\alpha=0$ and \SI{1160}{\nano\metre} for $\alpha=1$, against \SI{1132}{\nano\metre} in the data).

\begin{table}[H]
\centering
\caption{\textbf{Held-out interval widths at \SI{12}{\degreeCelsius}.} Central \SI{90}{\percent} intervals of $h(t+\tau)$ for releases starting in a $\pm\SI{35}{\nano\metre}$ window about the median gap, in nanometres.}
\label{tab:S_scores}
\begin{tabular}{lcc}
\toprule
Horizon & Data width & Model width ($\alpha=1$) \\
\midrule
\SI{10}{\milli\second}  & 141  & 128  \\
\SI{100}{\milli\second} & 366  & 398  \\
\SI{800}{\milli\second} & 1132 & 1160 \\
\bottomrule
\end{tabular}
\end{table}

\begin{table}[H]
\centering
\caption{\textbf{First-passage comparison from $h_0\approx\SI{280}{\nano\metre}$ to \SI{500}{\nano\metre}.} ``Data arrivals'' is the number of test releases that reach the threshold within the video; ``model fraction'' is the arrival probability predicted for the same observation window.}
\label{tab:S_fp}
\begin{tabular}{lccc}
\toprule
 & Data arrivals & Model fraction & Mean log-likelihood \\
\midrule
\SI{12}{\degreeCelsius}, $\alpha=0$ & 65/128 (\SI{51}{\percent}) & 0.56 & $-0.48\pm0.05$ \\
\SI{12}{\degreeCelsius}, $\alpha=1$ & 65/128 (\SI{51}{\percent}) & 0.67 & $-0.52\pm0.06$ \\
\SI{23}{\degreeCelsius}, $\alpha=0$ & 23/39 (\SI{59}{\percent})  & 0.56 & $-0.47\pm0.10$ \\
\SI{23}{\degreeCelsius}, $\alpha=1$ & 23/39 (\SI{59}{\percent})  & 0.69 & $-0.48\pm0.12$ \\
\bottomrule
\end{tabular}
\end{table}

\begin{table}[H]
\centering
\caption{\textbf{Prediction table at \SI{12}{\degreeCelsius}.} Mean displacement $\Delta h$ and CRPS on the test releases, in nanometres, for starting gaps between \SI{150}{\nano\metre} and \SI{500}{\nano\metre}. The uncertainty of the data is the standard error of a bootstrap over releases. ``Brown.'' is the local Brownian baseline. The last row has no empirical pair and reports the model only.}
\label{tab:S_full12}
\begin{tabular}{lcccccccc}
\toprule
 & Data & \multicolumn{3}{c}{Model $\Delta h$} & \multicolumn{4}{c}{CRPS} \\
\cmidrule(lr){3-5}\cmidrule(lr){6-9}
$\tau$ & $\Delta h$ & $\alpha=0$ & $\alpha=1/2$ & $\alpha=1$ & $\alpha=0$ & $\alpha=1/2$ & $\alpha=1$ & Brown. \\
\midrule
\SI{10}{\milli\second} & $0.7\pm1.2$ & $-0.2$ & $0.6$ & $1.3$ & $20.8$ & $20.8$ & $20.7$ & $20.4$ \\
\SI{20}{\milli\second} & $0.4\pm1.8$ & $0.4$ & $2.0$ & $3.5$ & $30.8$ & $30.7$ & $30.7$ & $30.5$ \\
\SI{50}{\milli\second} & $5.9\pm3.0$ & $2.3$ & $6.1$ & $9.9$ & $49.4$ & $49.2$ & $49.0$ & $49.2$ \\
\SI{100}{\milli\second} & $9.8\pm4.5$ & $5.0$ & $12$ & $20$ & $67.8$ & $67.3$ & $67.2$ & $67.4$ \\
\SI{200}{\milli\second} & $25\pm9$ & $11$ & $25$ & $40$ & $96$ & $95$ & $94$ & $96$ \\
\SI{400}{\milli\second} & $38\pm15$ & $23$ & $49$ & $75$ & $132$ & $131$ & $131$ & $132$ \\
\SI{600}{\milli\second} & $75\pm30$ & $34$ & $71$ & $108$ & $180$ & $178$ & $179$ & $182$ \\
\SI{800}{\milli\second} & $82\pm30$ & $46$ & $92$ & $138$ & $199$ & $197$ & $199$ & $201$ \\
\SI{1.00}{\second} & -- & $57$ & $112$ & $168$ & -- & -- & -- & -- \\
\bottomrule
\end{tabular}
\end{table}

\begin{table}[H]
\centering
\caption{\textbf{Prediction table at \SI{23}{\degreeCelsius}.} Mean displacement $\Delta h$ and CRPS on the test releases, in nanometres, for starting gaps between \SI{150}{\nano\metre} and \SI{500}{\nano\metre}. The uncertainty of the data is the standard error of a bootstrap over releases. ``Brown.'' is the local Brownian baseline. The last row has no empirical pair and reports the model only.}
\label{tab:S_full23}
\begin{tabular}{lcccccccc}
\toprule
 & Data & \multicolumn{3}{c}{Model $\Delta h$} & \multicolumn{4}{c}{CRPS} \\
\cmidrule(lr){3-5}\cmidrule(lr){6-9}
$\tau$ & $\Delta h$ & $\alpha=0$ & $\alpha=1/2$ & $\alpha=1$ & $\alpha=0$ & $\alpha=1/2$ & $\alpha=1$ & Brown. \\
\midrule
\SI{10}{\milli\second} & $2.3\pm2.1$ & $0.4$ & $1.2$ & $2.0$ & $20.9$ & $20.8$ & $20.7$ & $20.0$ \\
\SI{20}{\milli\second} & $1.8\pm2.9$ & $1.3$ & $2.9$ & $4.6$ & $29.5$ & $29.4$ & $29.3$ & $29.0$ \\
\SI{50}{\milli\second} & $13\pm5$ & $3.1$ & $7.2$ & $11$ & $48.7$ & $48.3$ & $48.0$ & $48.0$ \\
\SI{100}{\milli\second} & $21\pm9$ & $6.8$ & $15$ & $23$ & $65$ & $64$ & $64$ & $64$ \\
\SI{200}{\milli\second} & $41\pm17$ & $13$ & $28$ & $43$ & $86$ & $84$ & $83$ & $85$ \\
\SI{400}{\milli\second} & $75\pm29$ & $24$ & $53$ & $81$ & $124$ & $118$ & $116$ & $122$ \\
\SI{600}{\milli\second} & $94\pm51$ & $35$ & $74$ & $113$ & $164$ & $157$ & $155$ & $162$ \\
\SI{800}{\milli\second} & $98\pm60$ & $47$ & $95$ & $144$ & $199$ & $192$ & $190$ & $197$ \\
\SI{1.00}{\second} & -- & $54$ & $114$ & $175$ & -- & -- & -- & -- \\
\bottomrule
\end{tabular}
\end{table}

Averaged over the six starting gaps examined, from \SI{180}{\nano\metre} to \SI{430}{\nano\metre}, the mean first-passage log-likelihood is about $-0.33$ for each of the three conventions at both temperatures, so first passage to \SI{500}{\nano\metre}, which uses the same field, does not separate them.

\subsection{S3. Synthetic recovery of \texorpdfstring{$\alpha$}{alpha}}

For each true convention, 360 Euler--Maruyama trajectories were generated with the calibrated \SI{12}{\degreeCelsius} field, no tracker and a known $\alpha$, starting from gaps drawn uniformly between \SI{200}{\nano\metre} and \SI{320}{\nano\metre}. The first-passage log-likelihood recovers the correct convention by about \num{0.03} nats: for true $\alpha=0$ the log-likelihoods of the true and of the wrong convention are \num{-0.541} and \num{-0.575}; for true $\alpha=1$ they are \num{-0.522} and \num{-0.546}. The mean shifts at \SI{0.8}{\second} likewise follow the sign of the true convention. The method is therefore unbiased in direction, and the experimental videos are under-powered for the same test.

\subsection{S4. Estimators of spurious drift and \texorpdfstring{$\alpha$}{alpha}}

In each bin of $h$ we define
\begin{equation}
y=C-\frac{\Dperp}{\rho},\qquad
x=\widehat{D}_{\parallel}'+\frac{\widehat{D}_{\parallel}-\widehat{D}_{\perp}}{\rho},
\label{eq:S_xy}
\end{equation}
where hats denote the scaled hydrodynamic field, so that $y$ is the model-free spurious drift and $x$ the hydrodynamic divergence. The effective convention is the weighted regression through the origin
\begin{equation}
\widehat{\alpha}=\frac{\sum_b w_b\,x_b\,y_b}{\sum_b w_b\,x_b^2},\qquad
w_b=\frac{n_b}{\Dpar(h_b)},
\label{eq:S_alpha}
\end{equation}
with $n_b$ the number of increments in bin $b$, evaluated on bins with $\SI{100}{\nano\metre}\le h\le\SI{900}{\nano\metre}$ and sufficient counts. Uncertainties are the standard deviation of the same regression over bootstrap redraws of whole releases. $R^2$ is the weighted fraction of the variance of $y$ explained by $\widehat{\alpha}\,x$; a value near zero means that only the average of $y$, and not its shape in $h$, is informative.

\subsection{S5. Numerical details of the predictor}

For free evolution the predictor uses reflecting boundaries at $h=\SI{-550}{\nano\metre}$ and $h=\SI{2500}{\nano\metre}$, a grid spacing of \SI{8}{\nano\metre} and Crank--Nicolson steps of \SI{0.25}{\milli\second}. For first passage the grid reflects at $h=\SI{-400}{\nano\metre}$ and absorbs at \SI{500}{\nano\metre}. The training fraction is \num{0.70} and the random seed is 7. The horizons are 0.01, 0.02, 0.05, 0.1, 0.2, 0.4, 0.6, 0.8 and \SI{1.0}{\second}; the last one has no empirical pairs because the free window after frame 60 is shorter than \SI{1}{\second}.

\subsection{S6. Raw video against the laboratory file}

On the 117 releases at \SI{23}{\degreeCelsius} that exist both as a video and as a laboratory file, the projected gap was compared after removing a 41-frame moving average. The median correlation is \num{0.965} (tenth percentile \num{0.950}, ninetieth percentile \num{0.976}). Without that high-pass filter the median correlation is \num{-0.06}. The Brownian increments that enter $D$ at \SI{16.7}{\milli\second} are therefore the same measurement in the two records. The sub-hertz baseline is not, and it is large enough to move the spurious drift from \SI{86(42)}{\nano\metre\per\second} in the file to \SI{22(16)}{\nano\metre\per\second} on the raw frames. Both numbers are reported, and neither is promoted to a measurement of $\alpha$.

\subsection{S7. Why the defocus cue is not an axial coordinate}

\begin{figure}[H]
\centering
\includegraphics[width=0.56\linewidth]{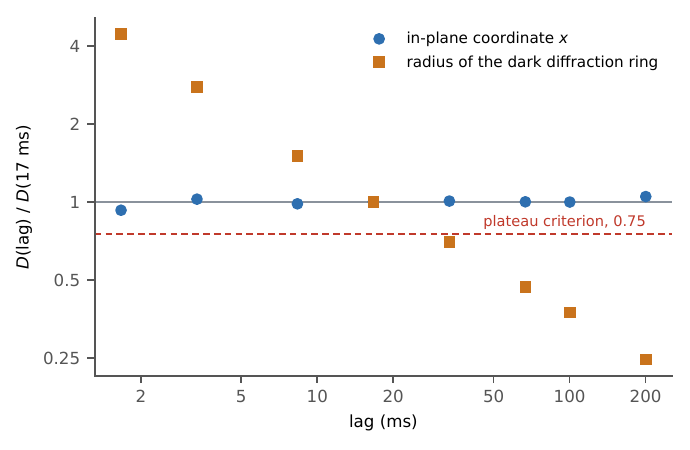}
\caption{\textbf{Plateau test of the defocus cue.} Apparent diffusivity against lag, divided by its value at \SI{17}{\milli\second}, for an in-plane coordinate and for the radius of the dark diffraction ring, both measured on the \SI{23}{\degreeCelsius} raw videos (608 releases). The in-plane coordinate stays near one; the ring radius falls through the plateau criterion (dashed line) and is not used.}
\label{fig:S_axial}
\end{figure}

The only axial cue available in the raw frames is the defocus of each sphere, measured here as the radius of its dark diffraction ring. Supplementary Fig.~\ref{fig:S_axial} applies to it the plateau test used for the in-plane coordinates, on the \SI{23}{\degreeCelsius} raw videos. The apparent diffusivity of $x$ stays within a few per cent of its value at \SI{17}{\milli\second} from \SI{3}{\milli\second} to \SI{200}{\milli\second}. That of the ring radius falls steadily, by a factor of about 18 over the same range; at a lag of \SI{100}{\milli\second} it is \num{0.37} of its value at \SI{17}{\milli\second}. Moreover, the ring radii of the two spheres differ by \num{0.52} pixels when the projected gap is positive and by \num{0.70} pixels when it is negative, against the $\sim$12 pixels of relative axial displacement that would be needed to explain an overlapping image by a real out-of-plane separation.

We identify four reasons for this failure; the first three are supported by the data and the fourth is a hypothesis.

\emph{The signal is dominated by noise.} For a coordinate that diffuses, the mean-square displacement grows linearly with lag and the apparent diffusivity is constant. A diffusivity that keeps falling with lag means that the mean-square displacement grows much more slowly than linearly, which is the signature of a stationary, noise-dominated variable: most of the frame-to-frame variation of the ring radius is measurement noise, not motion.

\emph{There is no calibration against a known displacement.} The ring radius was never recorded while a fixed sphere was displaced axially by a known amount, so there is no conversion from pixels of radius to nanometres of height, and no check that the relation is monotonic on both sides of focus.

\begin{figure}[H]
\centering
\includegraphics[width=0.92\linewidth]{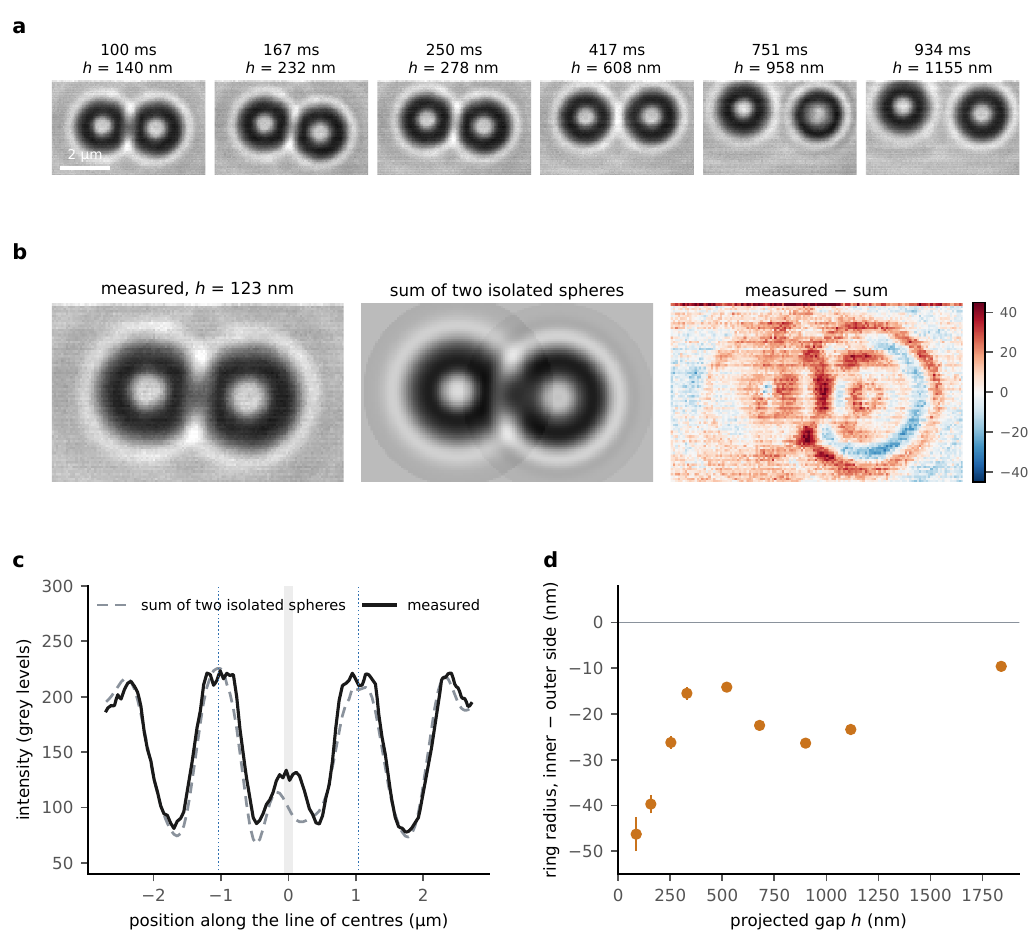}
\caption{\textbf{A release in raw frames, and the overlap of the diffraction rings at small gap.} (\textbf{a}) Six frames of one release at \SI{23}{\degreeCelsius}, with the time from the start of the record and the projected gap $h$; the pair diffuses and the gap opens. (\textbf{b}) Frame of smallest gap after frame 60 of the same release, the image expected if the two spheres did not affect each other (sum of two isolated-sphere images placed at the measured centres, built from the radial profile of each sphere at $h>\SI{1}{\micro\metre}$ on the side away from its neighbour), and their difference. (\textbf{c}) Intensity along the line of centres for the frame in (\textbf{b}): measured (solid) and sum of two isolated spheres (dashed). Dotted lines mark the centres and the grey band the gap. (\textbf{d}) Difference between the dark-ring radius measured on the side facing the neighbour and on the opposite side, against the gap; mean and standard error over 3224 sphere images from three releases.}
\label{fig:S_rings}
\end{figure}

\emph{At small gaps the diffraction patterns of the two spheres overlap.} Each sphere is imaged as a bright core, a dark ring and a bright outer halo (Supplementary Fig.~\ref{fig:S_rings}a). The halo extends beyond the geometrical edge of the sphere, so that when the gap is smaller than a few hundred nanometres the halo and the dark ring of one sphere fall on those of the other. The image between the spheres is then not the sum of two isolated-sphere images: at $h=\SI{123}{\nano\metre}$ a bright bridge appears on the line of centres where the sum of two isolated-sphere profiles predicts a minimum, and the difference between the measured frame and that sum reaches several tens of grey levels in a contrast of about 150 (Supplementary Fig.~\ref{fig:S_rings}b,c). The ring is consequently deformed on the side that faces the neighbour. Measured separately on the inner and on the outer side of each sphere, the ring radius is smaller on the inner side by \SI{46(4)}{\nano\metre} (\num{0.9} pixels) for $h<\SI{100}{\nano\metre}$, and by \SI{10(1)}{\nano\metre} for $h>\SI{1.3}{\micro\metre}$, although the decrease is not monotonic in between (Supplementary Fig.~\ref{fig:S_rings}d). A distortion of \num{0.9} pixels is larger than the difference of \num{0.5} to \num{0.7} pixels between the two spheres quoted above. The defocus cue therefore depends on the gap itself, and it is least reliable at the small gaps where the gradient of the diffusivity, and hence the spurious drift, is largest.

\emph{The immersion oil may add a temperature-dependent aberration.} As discussed in the main text, the oil is specified for operation near room temperature. Away from it the defocus response of the objective may change, which would make any calibration of the ring radius specific to one temperature. We have not tested this.

The deep-learning tracker of Helgadottir, Argun and Volpe~\cite{helgadottir2019} improves the in-plane centre under poor illumination but does not supply an axial calibration. Equation~\eqref{eq:sde} is therefore not extended to three dimensions.

\subsection{S8. Calibrated field and network}

\begin{figure}[H]
\centering
\includegraphics[width=0.82\linewidth]{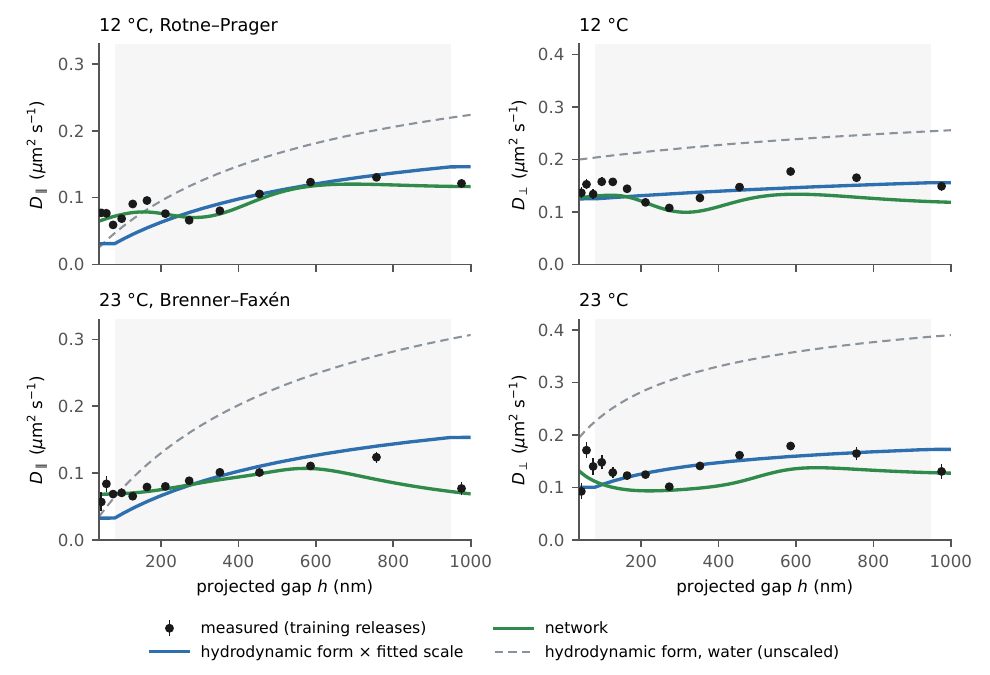}
\caption{\textbf{Calibrated diffusivity field and network.} $\Dpar(h)$ (left) and $\Dperp(h)$ (right) at \SI{12}{\degreeCelsius} (top) and \SI{23}{\degreeCelsius} (bottom). Points are the Kramers--Moyal profile measured on the training releases; the blue line is the retained hydrodynamic form multiplied by its fitted scale, which is the field used by the predictor; the dashed line is the same form for water, without scale; the green line is the network. The shaded band is the window used for the fit, \SIrange{80}{950}{\nano\metre}.}
\label{fig:S_net}
\end{figure}

Supplementary Table~\ref{tab:S_calib} lists the calibration of the hydrodynamic field on the training releases, and Supplementary Fig.~\ref{fig:S_net} compares that field with the network and with the measured profile. The scaled hydrodynamic form and the network agree on the level of the two diffusivities inside the fitted window; the network, which is not constrained to a hydrodynamic shape, departs from it at the smallest and largest gaps, where the samples are few.

\begin{table}[H]
\centering
\caption{\textbf{Calibration of the diffusivity field.} Number of training and test releases, hydrodynamic form retained, fitted scales and coefficients of determination for the two components.}
\label{tab:S_calib}
\begin{tabular}{lcccccc}
\toprule
Temperature & Training / test & Form & Scale $\parallel$ & Scale $\perp$ & $R^2_{\parallel}$ & $R^2_{\perp}$ \\
\midrule
\SI{12}{\degreeCelsius} & 338 / 146 & Rotne--Prager  & 0.66 & 0.61 & 0.31 & 0.22 \\
\SI{23}{\degreeCelsius} & 104 / 45  & Brenner--Faxén & 0.51 & 0.45 & 0.18 & 0.42 \\
\bottomrule
\end{tabular}
\end{table}

\end{document}